\documentclass
[12pt] {article}
\usepackage{amsmath}
\usepackage{amsfonts}
\usepackage{graphicx}

\newcommand{\bean}{\begin{eqnarray*}}
\newcommand{\eean}{\end{eqnarray*}}
\newcommand{\ed}{\end{document}}

\newcommand{\be}{\begin{equation}}
\newcommand{\ee}{\end{equation}}
\newcommand{\barr}{\begin{array}}
\newcommand{\earr}{\end{array}}
\newcommand{\bea}{\begin{eqnarray}}
\newcommand{\eea}{\end{eqnarray}}

\newcommand{\proof}{\medskip\noindent{\it Proof.}\quad }
\newcommand{\qed}{\hfill \fbox{}\medskip}

\begin{document}
\title{Explicit construction of the classical BRST charge for nonlinear algebras} 
\author{A.V.Bratchikov \\ Kuban State
Technological University,\\ 2 Moskovskaya Street, Krasnodar, 350072,
Russia
%\\ E-mail:bratchikov@kubstu.ru
} 
\date {March,\,2011} 
\maketitle

\begin{abstract}
We give an explicit formula for the Becchi-Rouet-Stora-Tyutin
(BRST) charge associated with Poisson superalgebras. To this end, we split the fundamental equation for the BRST charge into a pair of equations such that one of them is equivalent to the original one. We find the general solution to this equation. The solution possesses a graphical representation in terms of diagrams.
\end {abstract}
%\bigskip

%{\bf PAC} codes 03.65

%{\bf Keywords} hamiltonian systems with constraints.
%\bigskip

%\newpage
\section{Introduction}
The BRST symmetry \cite {BRS,T} plays an important role in quantization of gauge theories \cite {GT,HT}. It is generated by the BRST charge. 
If the quantum BRST charge exists it is essentially determined by the corresponding classical one.

The classical BRST charge is represented by a power series in ghosts. 
The first two terms of the series are well known in the general case. When constraints form a Lie algebra, these terms reproduce an exact BRST charge. 

The fundamental equation for the BRST charge is equivalent to a system of recurrent equations. In the case of general Poisson algebras there exists an algorithm for the construction of a solution to these equations \cite {FF}. 

For some  classes of quadratically nonlinear algebras the classical BRST charge was found in \cite {SSN,DH}.
Construction of the BRST charge for some boson Poisson algebras was investigated in \cite {BL}.  In the case of general quadratically nonlinear algebras the expression for the third order contribution in the ghost fields to the BRST charge was found. In \cite {ALRS}, the classical BRST charge for quadratically nonlinear superalgebras was discussed. 
For some  classes of  superalgebras the  BRST charge was constructed up to the fourth order in the ghost fields.

In this paper we derive an explicit expression of the classical BRST charge for nonlinear Poisson superalgebras.
We show that the system of equations for the classical BRST charge is equivalent to a smaller subsystem. 
Then we find the general solution to the subsystem. Expanding the solution in powers of the ghost fields one can find the BRST charge in an arbitrary order.

The paper is organized as follows.
In section 2, we introduce notations and represent the master equation
for the BRST charge in the form which is convenient for our
purposes. In section 3, we obtain  an explicit expression for the classical BRST charge. We show that 
the expression possesses a graphical representation in terms of diagrams.

In what follows Grassman parity and ghost number 
of a function $A$ are denoted by $\epsilon (A)$ and $\mbox{gh}(A),$ 
respectively.  

\section {Structure of the master equation for the BRST charge}
Let $G_\alpha, \alpha =1,\ldots,J,$ be the 
%initial 
first class constraints which satisfy the following Poisson brackets
\bean
%\label{p}
\{G_\alpha,G_\beta\}
=F_{\alpha \beta}(G),
\eean
where $F_{\alpha \beta}(G)$ is a polynomial in the $G$'s such that $F_{\alpha \beta}(0)=0.$ 
The constraints are supposed to be independent and of definite Grassmann parity $\epsilon_\alpha,$ 
$\epsilon(G_\alpha)=\epsilon_\alpha.$

Following the BRST method the ghost pair $( {\cal P}_\alpha,c^\alpha)$ is introduced for each constraint $G_\alpha:$
\bean 
%\label{U}
\{{\cal P}_\alpha, c^\beta, \}= \delta^{\alpha}_\beta, \quad  \{G_\alpha,c^\beta\}=\{{\cal P}_\alpha, G_\beta,  \}= 0,
\eean
$$\epsilon({\cal P}_\alpha)=\epsilon(c^\alpha)=\epsilon(G_\alpha)+1,$$ 
$$-\mbox {gh} ({\cal P}_\alpha) = \mbox {gh} (c^\alpha) =1.$$

Let ${\cal M}$ be the set of variables $(G^\alpha,{\cal P}_{\beta},c^{\gamma})
,$ and let 
${\cal V}=R[[{\cal M}]]$ be the ring of formal power series in the variables ${\cal M}.$

The BRST charge $\Omega\in {\cal V}$ is defined as a solution to the equation
\bea \label{o}
\{\Omega,\Omega\}=0, \quad \epsilon (\Omega)=1,\quad 
\mbox {gh} (\Omega)=1,
\eea
and the boundary conditions
\bea \label{bc}
\left.{\frac {\partial \Omega} {\partial {c^\alpha}}}\right|_{c=0} =G_\alpha.
\eea 
These equations are consistent \cite {H,BLT}. One can write
\bea \label{us}
\Omega= G_\alpha c^\alpha +M ,
\eea
where
\bean 
\label{sum}
M= \sum_{n= 2}^J\Omega^{(n)},\qquad  
\Omega^{(n)} \sim  {\cal P}^{n-1}c^n.
\eean

Substituting (\ref {us}) into (\ref {o})
one obtains 
%the recurrent equation
\bea \label{r}
\delta M +\frac 1 2  F + AM + \frac 1 2 \{M,M\}=0,
\eea
where
$$\delta = G_\alpha \frac { \partial_l} {\partial {\cal P}_\alpha},\qquad F= c^\alpha F_{\alpha \beta}(G) c^\beta,\qquad 
%\delta^2=0$$ 
%(acting from the left)
A=c^\alpha \{J_\alpha,\,.\, \}.$$

Let $
%\widehat 
N$ be the counting operator 
\bean 
%\label{r}
%\widehat 
N = G_\alpha \frac { \partial_l} {\partial {G}_\alpha}+{\cal P}_\alpha \frac {\partial_l} {\partial {\cal P}_\alpha}.
\eean
The space ${\cal V}$ splits as 
\bean 
%\label {deca}
{\cal V}= \bigoplus_{n\geq 0} {\cal V}_n
\eean
with  $
%\widehat 
N X=nX$ for $X\in {\cal V}_n.$
One easily verifies that  
$$
%\widehat 
N =\delta  \sigma+ \sigma \delta,\qquad 
\sigma= {\cal P}_\alpha \frac {\partial_l} {\partial G_\alpha}$$
$$
N\delta=\delta N , \qquad N\sigma=\sigma N.
$$

We define $N^+: {\cal V}\to{\cal V}$ by 
$$
N^+X =
\left \{
\begin {array}{rcl} 
\frac 1 {n}X,&\phantom {h} & X \in {\cal V}_n,\quad n>0;\\
0,\phantom {X}&\phantom {k} &  X \in {\cal V}_0.\\
\end{array}
\right.
$$
Then $\delta^{+}=\sigma N^+$ is a generalized inverse of $\delta$: 
\bean
% \label{d}
\delta\delta^{+}\delta=\delta,\qquad \delta^{+}\delta\delta^{+}=\delta^{+}
%,\quad (\delta^{+})^2=0
.
\eean

Let $\langle .\,,. \rangle: {\cal V}^2\to {\cal V}  $ be defined by
\begin{eqnarray*} \label {or}
\langle X_1,X_2 \rangle = -\frac 1 2 (I+\delta^+A)^{-1}\delta^{+}\left(\{ X_1,X_2 \}+\{ X_2,X_1 \}\right),
\end{eqnarray*} 
where $I$ is the identity map, and
$$(I+\delta^+A)^{-1}=\sum_{m\geq 0}(-1)^m(\delta^+A)^m.$$
\vspace{3mm}
\noindent

{\bf Lemma 1.} {
\it Eq. 
%(\ref {sum}),
(\ref {r}) is equivalent to
\bea 
\label{ome}
M = M_0 + \frac 1 2 \langle M,M \rangle,
\eea
%\bea \label{omef}(I-\delta\delta^+) D=0,\eea
where 
\bea 
\label{mo}
M_0 = (I+\delta^+A)^{-1}\left(Y-\frac 1 2 \delta^{+}F\right),
\eea 
and 
$Y\in {\cal V}$ is an arbitrary 
cocycle, 
$\delta Y=0,$ 
subject only to the restrictions 
\begin{eqnarray} 
\epsilon(Y)=1,\qquad \mbox{\rm gh}(Y)=1,\phantom{kkk}
%\quad
\nonumber 
\\
Y= \sum_{n= 2}^JY^{(n)},
\qquad  
Y^{(n)} \sim  
{\cal P}^{n-1}c^n.
\label{coc}
\end{eqnarray}
}

\proof 
In accordance with the decomposition 
$${\cal V}= 
{\cal V}_1 \oplus 
{\cal V}_2,$$
where 
$$ {\cal V}_1=P{\cal V},
\quad
{\cal V}_2= (I-P){\cal V},
\quad 
P=\delta\delta^+,$$
Eq.(\ref {r}) splits as  
\bea \label{c}
\delta M +\delta\delta^+ D=0,
\eea
\bea \label {d}
(I-\delta\delta^+) D=0,
\eea
where
$$D= \frac 1 2  F + AM + \frac 1 2 \{M,M\}.$$
From (\ref{c}) it follows that 
\bea M = Y - \delta^+ D,
\label {mu}
\eea
where the cocycle $Y\in {\cal V}$ satisfies (\ref {coc}). Eq. (\ref{mu}) can be rewritten in the form (\ref {ome}).   
%$\delta Y=0,$ $\epsilon(Y)=1,$ $\mbox{\rm gh}(Y)=1,$\begin{eqnarray*} Y= \sum_{n= 2}^JY^{(n)},\qquad  Y^{(n)} \sim  {\cal P}^{n-1}c^n.
%\end{eqnarray*}
One can show that 
$Y= \delta W,$ $W\in {\cal V}$ \cite {BLT}.

Eq. (\ref {ome}) can be iteratively solved as:
\bea 
\label{omey}
M = M_0 + \frac 1 2 \langle M_0,M_0 \rangle +
%\frac 1 2 \langle \langle M_{0},M_{0}\rangle, M_0 \rangle+
\ldots.
\eea
%(\ref {omey}) 
Using (\ref {omey}), we can write
$$
\Omega^{(n)}=\delta W^{(n)} +\widetilde \Omega^{(n)}\left(Y^{(2)}, Y^{(3)},\ldots, Y^{(n-1)}\right),\quad n\geq 2,    
$$
where $\delta W^{(n)}= Y^{(n)}.$
The arbitrariness of the solution for $\Omega^{(n)}$ is described by the transformation \cite {BT}
$$
\Omega^{(n)}\to \Omega^{(n)}+ \delta Z^{(n)}, \qquad Z\in {\cal V}.    
$$
This transformation is absorbed into a transitive group
transformation of the coboundary $\delta W^{(n)}$:
$$
\delta W^{(n)}\to \delta W^{(n)}+ \delta Z^{(n)}.    
$$
Therefore, $M$ (\ref {omey}) is the general solution to eq. (\ref {r}). It follows that Eq. (\ref {r}) is equivalent to (\ref {ome}). Eq. (\ref{d}) can be omitted.
\qed

In the next section, we obtain an explicit solution to Eq. (\ref {ome} ).

\section {Explicit expression for the BRST charge }

To solve eq. (\ref {ome}) we introduce the functions 
\begin{eqnarray*}\langle \ldots.  \rangle:
{\cal V }^m 
 \to {\cal V},\qquad m=1,2,\ldots,
  \end{eqnarray*}
which recursively 
defined by $$\langle X\rangle=X,$$
\begin{eqnarray} \label {u}
\langle X_1,\ldots ,X_m \rangle =
\frac 1 2 \sum_{r=1}^{m-1} \sum_{1\leq i_1<\ldots < i_r \leq m } \langle \langle X_{i_1},\ldots,X_{i_r} \rangle,
\langle X_1,\ldots,\widehat{X}_{i_1},\ldots,\widehat{X}_{i_r},\ldots,X_{m} \rangle 
 \rangle
\end{eqnarray} 
if $m=2,3,\ldots,$ where $\widehat{X}$ means that ${X}$ is omitted.

The following lemma is easily proved by induction.
\vspace{3mm}
\noindent

{\bf Lemma 2.} {\it $ \langle X_1,\ldots ,X_m \rangle$ is an $m-$ linear symmetric function.}
\vspace{3mm}
\noindent

For $m\geq 2,1\leq i,j\leq m,$ let  
$$P^m_{ij}: {\cal V}^m\to {\cal  V}^{m-1}$$  
be defined by
\begin{eqnarray*} \label {p}
P^m_{ij}( X_1,\ldots ,X_m ) = 
( \langle X_i,{X}_{j}\rangle ,X_1 ,\ldots,\widehat{X}_{i},\ldots,\widehat{X}_{j},\ldots,X_{m} ). 
\end{eqnarray*} 
If $X \in {\cal V}$ is given by  
\begin{eqnarray*} \label {v}  X = P^2_{12}P^3_{
i_{m-2}j_{m-2}
}\ldots P^{m-1}_
{i_2j_2}
P^m_{i_1j_1}
(X_1,\ldots ,X_m )
\end{eqnarray*} 
for some $ (i_1j_1),\ldots,(i_{m-2}j_{m-2}),$ we say that $X$ is a descendant of $(X_1,\ldots ,X_m ).
$
A descendant of $X\in {\cal V}$ is defined as $X.$

\begin{figure}\centering\includegraphics[width=2.0in]{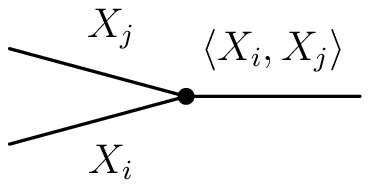}\begin{center} \caption{ 
%Diagram for 
$(X_i,X_j)\to \langle X_i,{X}_{j}\rangle.$ \label{fig1}}\end{center}\end{figure}

The function $\langle X_1,\ldots ,X_m \rangle$ can be described by diagrams. In these diagrams an element of  $\cal V$ is represented by the line segment \rule[3pt]{20pt}{0.5pt}\,\,. A product ${(X_i,X_j)\to \langle X_i,{X}_{j}\rangle}$ is represented by the vertex joining the line segments for $ X_i,{X}_{j}$ and $\langle X_i,{X}_{j}\rangle$ (see figure \ref{fig1}).

\vspace{3mm}
\noindent

{\bf Lemma 3.} \cite {B} {\it The function $\langle X_1,\ldots,X_m \rangle$ equals the sum of all the descendants of  $(X_1,\ldots ,X_m ).$}
\vspace{3mm}
\noindent
 
\begin{figure}
\centering
\includegraphics[width=3.5in]{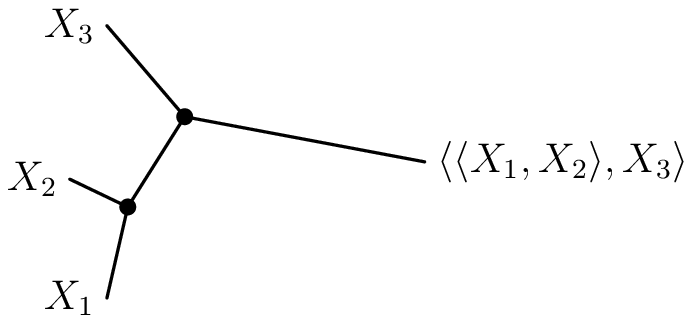}
\begin{center}
\caption {%Figure 2. 
Diagram for $\langle\langle X_1, X_2 \rangle, X_3\rangle.$ \label{fig2}}
\end{center}
\end{figure}

For example, 
%one gets
\begin{eqnarray*} \label {ord}
\langle X_1,X_2,X_{3}\rangle=
\langle \langle X_{1},X_{2}\rangle, X_3 \rangle+\langle \langle X_{1},X_{3}\rangle, X_2 \rangle+\langle  \langle X_2, X_3 \rangle,X_{1} \rangle. 
\end{eqnarray*} 
In figure \ref{fig2}, we show the diagram for $ \langle\langle X_1, X_2\rangle, X_3 \rangle.$

{\bf Lemma 4.} \cite {B} {\it A solution to Eq.(\ref {ome}) is given by
\begin{eqnarray} \label {orro}
M= \langle e^{M_0} \rangle,
\end{eqnarray}
where
$$\langle e^{M_0} \rangle =\sum_{m\geq 0} \frac {1} {m!}\langle M_0^m \rangle 
%\sum_{m=0}^\infty M_m
, \quad \langle M_0^r \rangle= 
\langle \underbrace {M_0,\ldots ,M_0}_
{
\text 
{{\it r} 
times
}
}
\rangle,\quad \langle M_0^0 \rangle=0.$$
} 
\vspace{3mm}
\noindent

Our previous results lead to the following theorem.
%We denote $$M_m=\frac {1} {(m+1)!} \langle M_0^{(m+1)}\rangle
%,\qquad  m=0,1,\ldots.$$
\vspace{3mm}
\noindent

{\bf Theorem.} 
{\it  
The general solution to Eqs. (\ref {o}),(\ref {bc}) is given by 
\bea \label {mi} \Omega= G_\alpha c^\alpha+\sum_{m\geq 1}\frac {1} {m!}\langle M_0^m \rangle
.
\eea}

\vspace{3mm}
\noindent

From (\ref{mo}) it follows immediately that 
$M_0= O(c^2).$ 
Using (\ref{u}) and the induction method, one can show that
\bean \label {m}
\langle M_0^m \rangle = O(c^{m+1}).
\eean
Hence, in the case of bosonic constraints, $\epsilon(G_\alpha)=0,\epsilon(c^\alpha)=1,$ eq. (\ref {mi}) takes the form
$$\Omega= G_\alpha c^\alpha+\sum_{m=1}^{J-1}\frac {1} {m!}\langle M_0^m \rangle
.$$

For example, using 
lemma
1, 
we get
$$\Omega=G_\alpha c^\alpha+\sum_{m=1}^5 \frac {1} {m!}\langle M_0^m\rangle +O(c^7), $$
where
\begin{eqnarray*}
\frac {1} {3!}\langle M_0^3 \rangle=
 \frac 1 2 \langle \langle M_{0},M_{0}\rangle, M_0 \rangle,\phantom{\frac 1 4 
\langle
\langle\langle A_{0},S_{0}\rangle,D_0 
\rangle 
\langle 
G_{0},F_{0}
\rangle,
\rangle+k}
\,\,\,\,\,\,\,\,\,\,\,\,\,\,\,\,\,\,\,\,\,\,\,\,\,\,\,\,\,\,\,\,\,\,\,\,\,\,\,\,\,\,\,\,
\end{eqnarray*}  
\begin{eqnarray*} 
\frac {1} {4!}\langle M_0^4 \rangle
=  \frac 1 2 \langle \langle \langle M_{0},M_{0}\rangle,M_{0}\rangle, M_{0}\rangle+\frac 1 8 \langle \langle M_{0},M_{0}\rangle, \langle M_{0}, M_{0}\rangle\rangle,\phantom{ 
\langle
\langle\langle A_{0},S_{0}\rangle,D_0 
\rangle 
}
\end{eqnarray*}  
\begin{eqnarray*} 
\frac {1} {5!}\langle M_0^5 \rangle
= 
\frac 1 2
\langle \langle
\langle \langle 
M_{0},M_{0}\rangle,
M_{0}\rangle,M_{0}\rangle, M_{0}\rangle+
\frac 1 4 
\langle
\langle\langle M_{0},M_{0}\rangle,M_0 
\rangle, 
\langle 
M_{0},M_{0}
\rangle
\rangle+\\
 +\frac 1 8 
\langle
\langle 
\langle 
M_{0},M_{0}
\rangle,
\langle 
M_{0},M_{0}
\rangle \rangle
,M_{0}
\rangle.
\phantom{\frac 1 4 
\langle
\langle\langle A_{0},S_{0}\rangle,D_0 
\rangle 
\langle 
G_{0},F_{0}
\rangle,
\rangle+kl\,\,\,}
\end{eqnarray*}

%\newpage

%\bigskip

\end{document}